\documentclass[prd,twocolumn,showpacs,floatfix,amsmath,amssymb,floatfix]{revtex4}
\usepackage{graphicx,color,dcolumn,booktabs,bm}
\usepackage{longtable,lscape}
\usepackage{txfonts}
\usepackage{overpic}
\usepackage{amssymb}
\usepackage{indentfirst}
\usepackage{feynmf}   
\usepackage{slashed}  
\usepackage{cases}
\usepackage{color}
\usepackage{multirow}
\usepackage{graphicx,color,dcolumn,booktabs,bm}
\usepackage{epsfig,dsfont,amssymb,amsmath,amsfonts,amsbsy,mathrsfs}
\usepackage{subfigure}

\graphicspath{{Figures/}} %

\begin{document}

\title{The leptonic CP  phase is determined by an equation involving the PMNS matrix elements }

\vspace{1cm}

\author{ Hong-Wei Ke$^1$\footnote{khw020056@hotmail.com, Corresponding author.}, Jia-Hui Zhou and
          Xue-Qian Li$^2$\footnote{lixq@nankai.edu.cn} }

\affiliation{  \\$^{1}$  School of Science, Tianjin University,
Tianjin 300072, China
 \\$^{2}$ School of Physics, Nankai University, Tianjin 300071, China
}

\vspace{12cm}

\begin{abstract}
Several approximate equalities among the matrix elements of CKM
and PMNS imply that hidden symmetries may exist and be common for
both quark  and neutrino sectors. The CP phase of the CKM matrix
($\delta_{\rm CKM}$) is involved in these equalities and can be
investigated when these equalities turn into several equations. As
we substitute those experimentally measured values of the three
mixing angles into the equations for quarks, it is noted that one
of the equations which holds exactly has a solution $\delta_{\rm
CKM}=68.95^\circ$. That value accords with
$(69.1^{+2.02}_{-3.85})^\circ$ determined from available data.
Generalizing the scenario to the lepton sector, the same equality
determines the leptonic CP phase $\delta_{\rm PMNS}$ to be $
276.10^\circ$. Thus we predict the value of $\delta_{\rm PMNS}$
from the  equation. So far there is no direct measurement on
$\delta_{\rm PMNS}$ yet, but a recent analysis based on the
neutrino oscillation data prefers the phase close to $270^\circ$.


\end{abstract}

\maketitle

\section{Introduction}
Following recognition of quark mixing and determining the non-zero
CP phase in the quark sector, with longtime accumulation of data
on solar\cite{Davis:1968cp,Ahmad:2002jz,Aharmim:2005gt},
atmosphere\cite{Fukuda:1998mi,Ashie:2004mr},
accelerator\cite{Ahn:2002up,Michael:2006rx,Adamson:2008zt} and
reactor neutrino
experiments\cite{Eguchi:2002dm,Abe:2008aa,Abe:2011fz,Abe:2012tg,An:2012eh,Ahn:2012nd},
the mixing among different neutrino flavors is confirmed. {Even 5
to 10 years before the first measurement on the solar neutrino
flux had been made, the neutrino
mixing\cite{Pontecorvo:1957cp,Maki:1962mu} was proposed and with
the picture\cite{Pontecorvo:1967fh}, the solar neutrino flux
deficit measured fifty years ago can be naturally interpreted.}
Since the lepton flavor eigenstates do not match with the mass
eigenstates a unitary transformation matrix i.e.
Pontecorvo-Maki-Nakawaga-Sakata (PMNS) matrix was introduced to
bridge them. The $3\times 3$ matrix possesses three mixing angles
and a CP  phase which may also be non-zero. From then on, to
determine the mixing parameters and CP phase composes the most
important task for both experimentalists and theorists of this
field. Even though through hard work the three mixing angles have
been determined with a certain accuracy, the leptonic CP phase is
still not clear. Along with the experimental search for the CP
pase, one might expect to predict it from the theory aspect.

In our early works\cite{Ke:2014gda,Ke:2015sba} several approximate
equalities among the matrix elements of Cabibbo-Kobayashi-Maskawa
(CKM)\cite{Cabibbo:1963yz,Kobayashi:1973fv} were found based on
investigating relations among different parametrization schemes of
the mixing matrix (CKM and PMNS). The equalities may be induced by
a hidden symmetry. It is noted that those equalities also hold for
the lepton sector, even though the present data on neutrinos are
not as accurate as for the quark sector.  Obviously those
equalities may help to reduce the number of free parameters of the
matices, i.e. the degree of freedom  is reduced due to existence
of  the hidden symmetry. For that we regard these equalities to
several equations. Supposing these equations to be exact, by
inputting several CKM matrix elements\cite{PDG14,Kang:2013jaa}
into the equations, other elements are obtained, and it is found
that the results well coincide with the available data  within the
error tolerance range. In the three-generation fermion framework,
there are only four free parameters in the $3\times 3$ mixing
matrix i.e. the three mixing angles and one CP phase, thus we
expect that due to a hidden symmetry one or more parameters could
be determined by the other parameters via the equations. In this
work, we are able to determine the CP phase of CKM from the
equations and then generalize this scenario to the lepton sector.
Meanwhile, we would testify which equality holds more precisely or
in other words, to judge their approximation degrees. The results
help us to conclude whether the hidden symmetry is complete or
somehow slightly broken.  In this paper we will use simplified
forms of these equations which were given in our earlier works and
investigate the values of the leptonic CP phase by these
equations.

\section{Equalities among the matrix elements and the CP phase}

Mixing among different flavors of quarks (neutrinos) via the CKM
(PMNS) matrix has been firmly recognized. The Lagrangian of the
weak interaction reads
\begin{eqnarray} \label{Lag}
\mathcal{L}=\frac{g}{\sqrt{2}}\bar {U}_L\gamma^\mu V_{\rm CKM} D_L
W^+_{\mu}+\frac{g}{\sqrt{2}}\bar E_L \gamma^\mu V_{\rm PMNS} N_L
W^-_{\mu}+h.c.,
\end{eqnarray}
where $U_L=(u_L, c_L, t_L)^T$,  $D_L=(d_L, s_L, b_L)^T$,
$E_L=(e_L, \mu_L, \tau_L)^T$and  $N_L=(\nu_1, \nu_2, \nu_3)^T$.
$V_{\rm CKM}$ and $V_{\rm PMNS}$ are the CKM  and PMNS matrices
respectively.  The $3\times 3$ mixing matrix $V$ is written as
    \begin{equation}\label{M1}
      V=\left(\begin{array}{ccc}
        V_{11} &V_{12} &V_{13} \\
         V_{21} &   V_{22} &  V_{23}\\
          V_{31} & V_{32} & V_{33}
      \end{array}\right).
  \end{equation}

Generally, for a $3\times 3$ unitary matrix there are four
independent parameters, namely three mixing angles and one CP phase.
There can be various schemes to parameterize the matrix which are
summarized in Ref.\cite{Zhang:2012pv}. The standard parametrization
is shown as\cite{PDG14}
 \begin{equation}\label{M2}
      V=\left(\begin{array}{ccc}
        c_{12}c_{13} & s_{12}c_{13} & s_{13}\\
       -c_{12}s_{23}s_{13} -s_{12}c_{23}e^{i \delta} & -s_{12}s_{23}s_{13}+c_{12}c_{23}e^{i\delta} & s_{23}c_{13}\\
      -c_{12}s_{23}s_{13} + s_{12}s_{23}e^{i\delta} & -s_{12}s_{23}s_{13}-c_{12}s_{23}e^{i\delta} & c_{23}c_{13}
      \end{array}\right).
\end{equation}
Here $s_{jk}$ and $c_{jk}$ denote $\sin\theta_{jk}$ and
$\cos\theta_{jk}$ with $j,k=1,2,3$. $\delta$ is the CP phase and
$\delta_{\rm CKM}$ and $\delta_{\rm PMNS}$ will be used for CKM and
PMNS respectively.

\begin{figure*}
        \centering
        \subfigure[~Eq. (9)\,(CKM)]{
          \includegraphics[width=5.8cm]{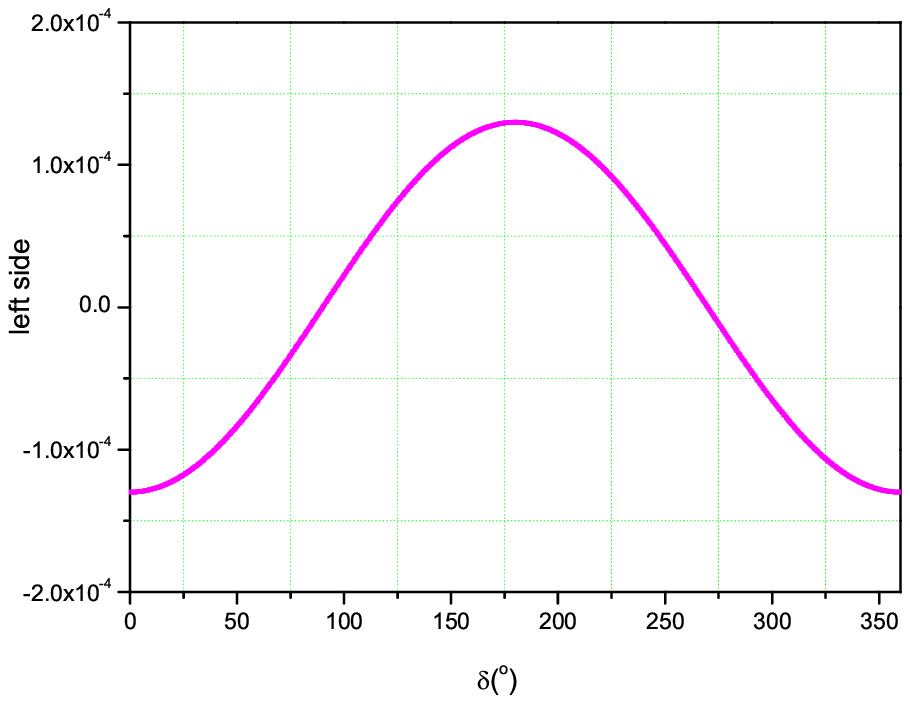}}
        \subfigure[~Eq. (10)\,(CKM)]{
          \includegraphics[width=5.8cm]{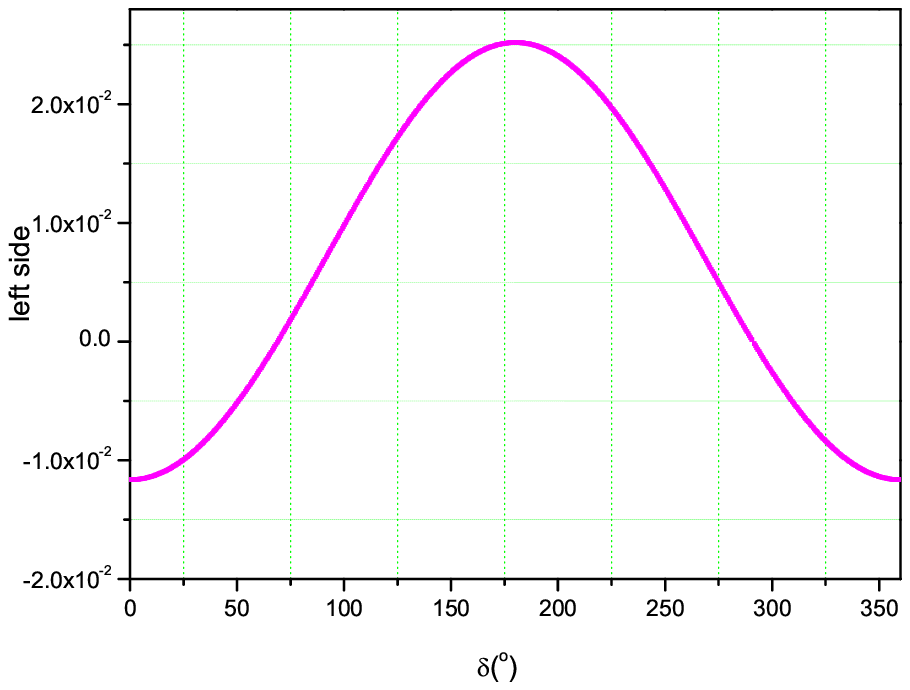}}
        \subfigure[~Eq. (11)\,(CKM)]{
          \includegraphics[width=5.8cm]{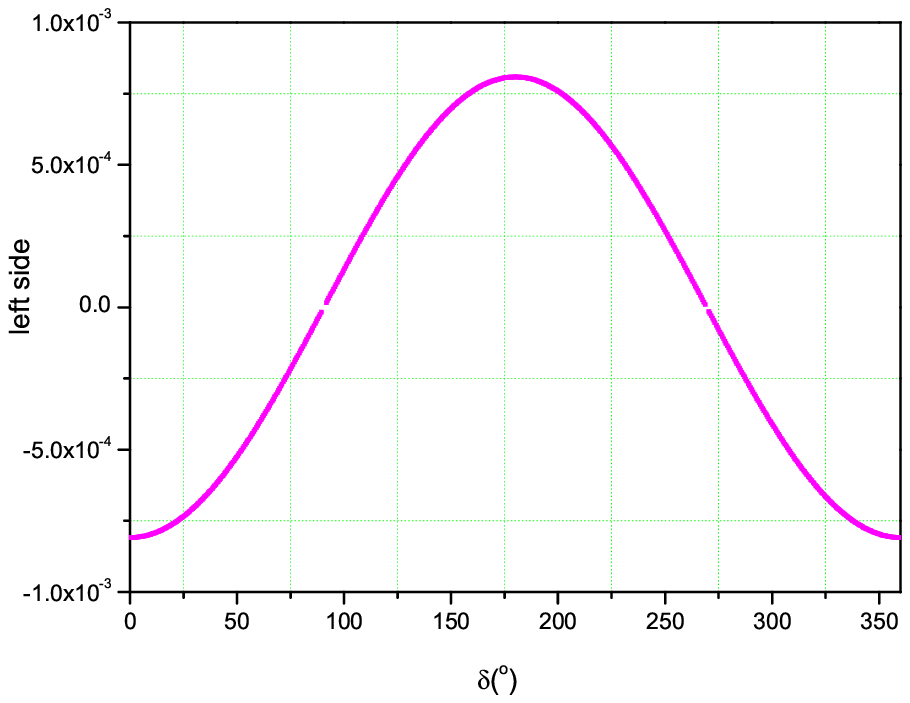}}
        \subfigure[~Eq. (12)\,(CKM)]{
          \includegraphics[width=5.8cm]{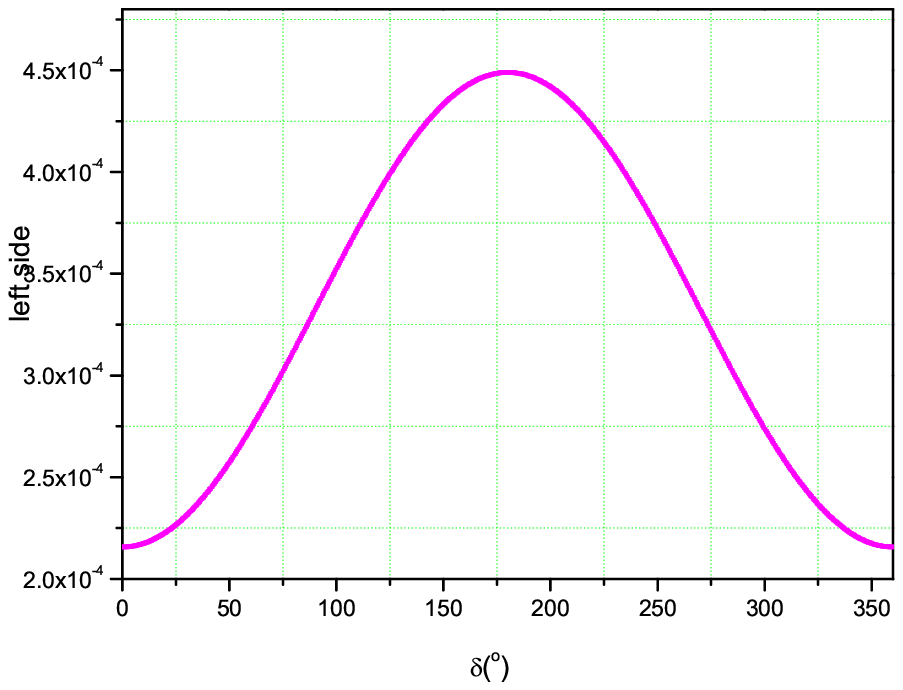}}
        \subfigure[~Eq. (13)\,(CKM)]{
          \includegraphics[width=5.8cm]{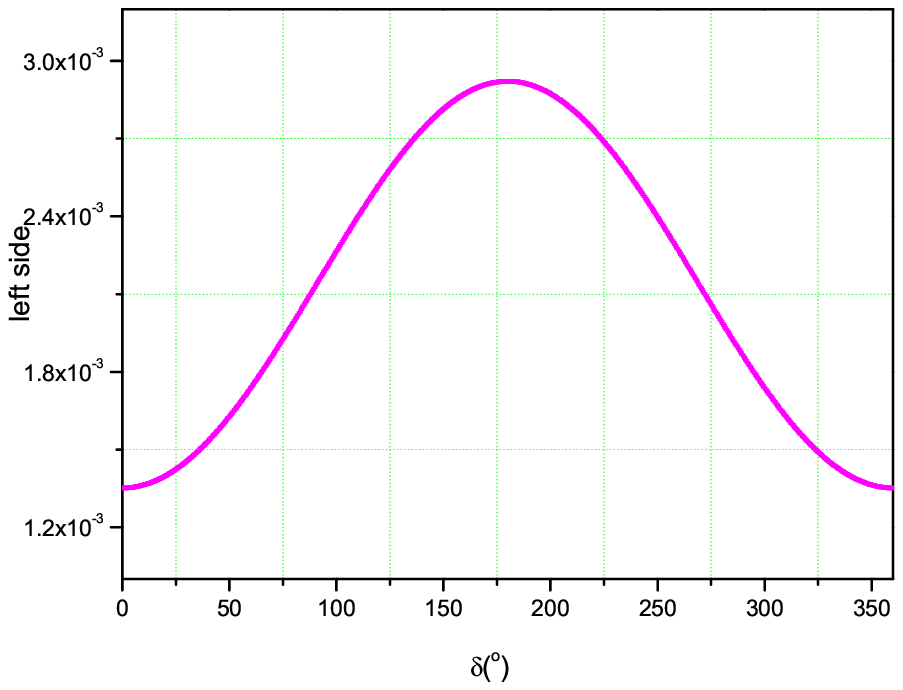}}
           \subfigure[~Eq. (10)\,(PMNS)]{
          \includegraphics[width=5.8cm]{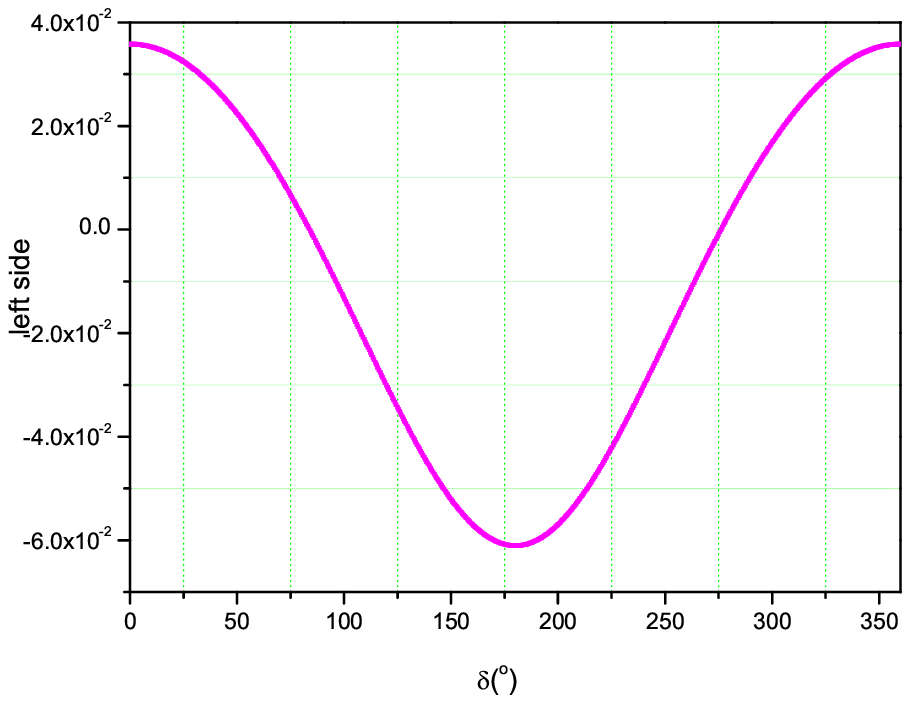}}
 \caption{The dependence of the left sides of Eq. (4)-Eq. (8) on the CP phase $\delta$}
        \label{fig4}
    \end{figure*}

In our previous work \cite{Ke:2014gda,Ke:2015sba} definite
relations among the values of $\sin\delta_i$ where $\delta_i$'s
are the CP phase in the nine different parametrization schemes
were found, then we turned these relations into several
approximate equalities among the matrix elements of CKM (or PMNS).

These equalities can be then rewritten to more succinct forms
\begin{eqnarray} \label{rl9}
 \frac{{|V_{21}|} {|V_{22}|}
}{
   1-{|V_{23}|}^2}-\frac{{|V_{11}|} {|V_{12}|}
   }{
    {1-|V_{13}|}^2}\approx0 ,
    \end{eqnarray}
    \begin{eqnarray} \label{rl10}
   \frac{{|V_{11}|}  {|V_{12}|} {|V_{21}|}}{
   1-{|V_{11}|}^2}
  -\frac{{|V_{23}|} {|V_{32}|} {|V_{33}|}  }{
   1-{|V_{33}|}^2}\approx0 ,
   \end{eqnarray}
   \begin{eqnarray} \label{rl11}
 \frac{{|V_{23}|}  {|V_{33}|}}{
   1-{|V_{13}|}^2}-\frac{ {|V_{22}|}
   {|V_{32}|}}{
   {1-|V_{12}|}^2} \approx0,
   \end{eqnarray}
   \begin{eqnarray} \label{rl12}
 \frac{{|V_{12}|} {|V_{22}|}}{
   1-{|V_{32}|}^2}-
   \frac{{|V_{11}|} {|V_{21}|} }{1-{|V_{31}|}^2
  }\approx0 ,
  \end{eqnarray}
  \begin{eqnarray} \label{rl13}
 \frac{ {|V_{22}|}  {|V_{23}|}}{
   {1-|V_{21}|}^2}-\frac{ {|V_{32}|}  {|V_{33}|}}{1-{|V_{31}|}^2
   }\approx0
  ,
\end{eqnarray}
{where normalization conditions  such as
${|V_{13}|}^2+{|V_{23}|}^2+{|V_{33}|}^2=1$ were used and
$|V_{jk}|\,(j,k=1,2,3)$ are the modules of the corresponding
matrix elements.} Apparently Eqs. (\ref{rl9}) and (\ref{rl13})
display relations among the matrix elements of the two adjacent
rows and Eqs. (\ref{rl11}) and (\ref{rl12}) demonstrate relations
among the matrix elements of the two adjacent columns. Eq.
(\ref{rl10}) corresponds to the relations among the six matrix
elements ${|V_{11}|}, {|V_{12}|}, {|V_{21}|},{|V_{23}|},
{|V_{32}|}$ and $ {|V_{33}|}$. These equalities reveal existence
of some underlying symmetries which determine them.

Now let us suppose the matrix elements to be variables, then these equalities
would turn into several equations by which one may expect to gain solutions. In this letter we will
further investigate the CP phases of CKM and PMNS in terms of the
corresponding equations. Since there are abundant data on the
hadron decays, the parameters in the CKM matrix are well
determined. One can use these values to judge which equality holds
with higher precision.

Using the expressions of matrix elements in Eq. (\ref{M2}), Eqs.
(\ref{rl9})-(\ref{rl13}) are transformed into the forms directly
related to the three mixing angles and CP  phase. If one inputs the
values of three mixing angles he can study the dependence of the
left sides of Eqs. (\ref{rl9})-(\ref{rl13}) on the CP phase. For
quark sector the dependence is depicted in Fig. 1(a)-(e) with
$\theta_{12}=13.023^\circ, \theta_{23}=2.360^\circ,
\theta_{13}=0.201^\circ$\cite{Zhang:2012pv}. From those diagrams,
one finds:

1. All the left sides of the five equations vary within very small
ranges for different $\delta_{\rm CKM}$ values.

2. At the two sultions the first three equations hold
exactly. We list the corresponding values in Table
\ref{tab:value}.

3. The last two
equations do not possess solutions which are reasonable, namely the values of the left sides
of the equations are
close to 0 when $\delta_{\rm CKM}=0^\circ$ or $360^\circ$ that obviously contradict to the suggested
data.

4. The left side of Eq. (\ref{rl10}) is sensitive to the change of
$\delta_{\rm CKM}$.

5. The solution $68.95^\circ$ obtained from Eq. (\ref{rl10}) is
accordant with the data
$(69.1^{+2.02}_{-3.85})^\circ$\cite{Zhang:2012pv} within the error
tolerance i.e. the equality offers proper information on the hadronic CP
phase for the quark sector.
\begin{table}
\caption{The values of the CP  phase which make the five equations
to hold more accurately.} \label{tab:value}
\begin{ruledtabular}
\begin{tabular}{ccc|cc}
 Eq. &$\delta_{\rm CKM}$(1)  &
$\delta_{\rm CKM}$(2)  &$\delta_{\rm PMNS}$(1)  & $\delta_{\rm
PMNS}$(2)
\\\hline
(\ref{rl9})&90.01 $^\circ$  &269.98$^\circ$ &93.21$^\circ$   &266.79$^\circ$ \\
(\ref{rl10})&68.95 $^\circ$  &291.05$^\circ$  &83.90$^\circ$&276.10$^\circ$\\
(\ref{rl11})&90.57$^\circ$   &269.43$^\circ$&90.74 $^\circ$  &269.26$^\circ$ \\
(\ref{rl12})&0 $^\circ$ &360$^\circ$&0 $^\circ$ &360$^\circ$ \\
(\ref{rl13})&0 $^\circ$ &360$^\circ$&0 $^\circ$ &360$^\circ$ \\
\end{tabular}
\end{ruledtabular}
\end{table}

Apparently Eq. (\ref{rl10}) holds with the highest accuracy
whereas Eqs. (\ref{rl9}) and (\ref{rl11}) are not so precise, but
the approximation degree is still appreciable, then for expressions of
(\ref{rl12}) and (\ref{rl13}) equality is almost totally lost to a
not-allowed degree. Thus it means that the equality degree of Eq.
(\ref{rl9})  through Eq. (\ref{rl13})  could be different as the
underlying symmetry is broken, therefore not all of them can be
applied to determine the CP phase by inputting the mixing angles.
We decide that Eq.(\ref{rl10}) would be the best choice.

So far there is no direct measurement on the leptonic CP phase yet,
naturally one may expect to gain information on it from future experiments. It is
known that $\delta_{\rm PMNS}$ can induce CP violation effects in
neutrino oscillation experiments. Better than nothing, at present, indirect
analyses\cite{Abe:2013hdq,Gonzalez-Garcia:2014bfa} based on the
neutrino experiments suggest that the leptonic CP phase $\delta_{\rm PMNS}$
is close to $270^\circ$.
Since the Eq. (\ref{rl10}) predicts a more precise value of $\delta_{\rm CKM}$,
we have all reason to expect that  it is eligible to be applied for determining $\delta_{\rm
PMNS}$. In analog to the quark case as inputting the central values of
$\theta_{12}=33.71^\circ, \theta_{23}=41.38^\circ,
\theta_{13}=8.80^\circ$ (for normal ordering of neutrino masses) one
can fix $\delta_{\rm PMNS}=83.90^\circ$ or $276.10^\circ$ as Eq.
(\ref{rl10}) is supposed to hold exactly.
For inverted ordering of neutrino masses the central value of
$\theta_{13}$ is $8.91^\circ$ and then the predicted value of
$\delta_{\rm PMNS}$ would change only slightly. Apparently the
theoretical prediction $\delta_{\rm PMNS}\sim 276.10^\circ$ is
consistent with the results given in Ref.
\cite{Abe:2013hdq,Gonzalez-Garcia:2014bfa}. If the result is
confirmed by the future measurements one would find that the sum
of $\delta_{\rm CKM}$ and $\delta_{\rm PMNS}$ is close to
$360^\circ$ i.e. $\delta_{\rm CKM}$ adopts the left side value of
Fig. \ref{fig4}(b) whereas $\delta_{\rm PMNS}$ takes the right
side value (f). The mutual complementarity
between $\delta_{\rm PMNS}$ and $\delta_{\rm CKM}$ may impose a constrant
condition to exclude the solution of $83.90^\circ$. From Tab.
\ref{tab:value} one can note that the solutions obtained from Eqs.
(\ref{rl9}) and (\ref{rl11}) are also close to $270^\circ$, even though
in the quark sector the CP phase determined by their counterpart equations apparently deviates
from data.

\section{SUMMARY}

In our previous paper several approximate equalities among the
elements of CKM (PMNS) were derived. In this letter we try to
simplify them and those simple and succinct relations imply
existence of hidden symmetries  in both the quark and neutrino
sectors even though they may be approximate or slightly broken.
Using the equations we investigate the CP phase of CKM and PMNS
matrices. Supposing these equations hold exactly, as the three
mixing angles of quarks are inputs one only obtains an accurate
$\delta_{\rm CKM}$ from Eq. (\ref{rl10}) whereas the others
equations fail to provide precise information about the CP phase,
even though the solutions of Eq. (\ref{rl9}) and Eq. (\ref{rl11})
are still plausibly close to the data. By solving Eq. (\ref{rl10})
with inputting the the measured values of the neutrino mixing
angles we obtain the leptonic CP phase $\delta_{\rm PMNS}$ to be
$83.90^\circ$ or $276.10^\circ$. Since at present there is no
direct experimental measurement on $\delta_{\rm PMNS}$ of PMNS
yet, we need to determine which value of the two would be that
chosen by the nature. The mutual complementarity between quark and
lepton sectors help to pin it down. Moreover, indirect analyses
made by experimentalists indicate that the phase is close to
$270^\circ$. Thus the solution $\delta_{\rm PMNS}=276.10^\circ$ is
consistent with the recent analyses on neutrino experiment.

It is well known that the  $3\times 3$ unitary matrix for fermion
mixing can have three mixing angles and one CP phase. Generally,
except the unitary requirement, it seems that there are no other
constraints on them, so the four parameters are independent.
Namely, one cannot derive any of them from the other three.
However, our analysis clearly shows that from the equations which
were derived in our earlier works the CP phase can be obtained
from the three mixing angles whose values are experimentally
measured. It implies that there is a hidden symmetry which may
associate the four parameters together. Even though we do not know
what the hidden symmetry is, we notice its existence. The hidden
symmetry determines those equalities which we discovered from
phenomenology. Therefore, precise data of the three mixing angles
would determine the CP phase except there exist a $\delta$ and
$2\pi-\delta$ degeneracy. That is the common sense.  Moreover, the
association between quark and lepton sectors is exposed and this
generalization of those equalities from quark sector to leptonic
one is by no means trivial. In particular, we use the same
equality which holds for both quarks and leptons to obtain the CP
phase as $276.10^{\circ}$ or $83.90^{\circ}$ for neutrinos which
has not well been experimentally measured yet but is somehow
determined by the oscillation data, and our theoretical prediction
is consistent with  this value. In this sense, the degeneracy of
$\delta$ and $2\pi-\delta$ is lifted by the mutual complementarity
between quark and lepton sectors as $\delta_{PMNS}=270^{\circ}$
being uniquely fixed. In this work we only expose this fact from
the phenomenological aspect, but it is still mysterious if all the
equalities indeed originate from a hidden symmetry. If so, does
the phenomenon that some of the equalities are only approximately
holding, mean breaking of the symmetry or there exists a mechanism
which causes deviation from exact holding? The profound source is
worth of further exploration.

\section*{Acknowledgement}

This work is supported by the National Natural Science Foundation
of China (NNSFC) under the contract No. 11375128 and 11135009.

\appendix


\end{document}